# A MEMS-based optical scanning system for precise, high-speed neural interfacing

Cem Yalcin, *Student Member, IEEE*, Nathan Tessema Ersaro, *Student Member, IEEE*, M. Meraj Ghanbari, *Student Member, IEEE*, George Bocchetti, *Member, IEEE*, Sina Faraji Alamouti, *Student Member, IEEE*, Nick Antipa, *Member, IEEE*, Daniel Lopez, Nicolas C. Pégard, Laura Waller, Rikky Muller, *Senior Member, IEEE*

*Abstract*— **Optical scanning is a prevalent technique for optical neural interfaces where light delivery with high spatial and temporal precision is desired. However, due to the sequential nature of point-scanning techniques, the settling time of optical modulators is a major bottleneck for throughput and limits random-access targeting capabilities. While fast lateral scanners exist, commercially available varifocal elements are constrained to >3ms settling times, limiting the speed of the overall system to hundreds of Hz. Faster focusing methods exist but cannot combine fast operation and dwelling capability with electrical and optical efficiency. Here, we present a varifocal mirror comprised of an array of piston-motion MEMS micromirrors and a custom driver ASIC, offering fast operation with dwelling capability while maintaining high diffraction efficiency. The ASIC features a reconfigurable nonlinear DAC to simultaneously compensate for the built-in nonlinearity of electrostatic actuators and the global process variations in MEMS mirrors. Experimental results demonstrate a wide continuous sweeping range that spans 22 distinctly resolvable depth planes with refresh rates greater than 12 kHz.**

## I. INTRODUCTION

ALL-OPTICAL neural interfaces are a promising class of tools for neuroscience research that enable simultaneous monitoring and manipulation of neuronal activity with light. New devices specifically designed to optically address neurons are now within reach thanks to recent advances in imaging and stimulation capabilities [1]. On the imaging front, emerging genetically encoded voltage indicators (GEVIs) can encode single cell potentials down to mV levels into fluorescence signals, with response times of hundreds of microseconds to milliseconds [2]. For optical stimulation, neurons can be virally or genetically modified to express light-sensitive proteins (opsins) that excite or inhibit neural activity in response to light at specific wavelengths [3]. State-of-the-art opsins reliably respond with exposure times on the order of a few milliseconds and with sub-millisecond jitter performance [4]. As the spatio-temporal resolution of neural imaging and stimulation modalities advance, accurate and high-speed delivery of excitation light for the interrogation or modulation of the neural activity is becoming the main bottleneck limiting the performance of all-optical neural interfaces.

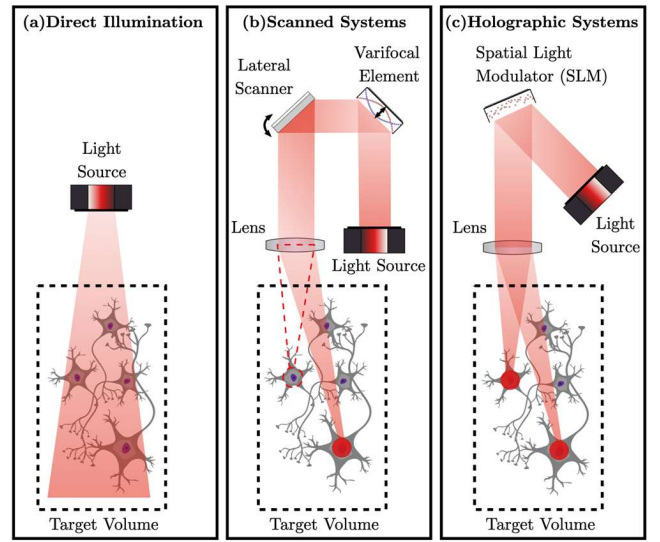

Fig. 1: Simplified diagrams of light delivery systems for all-optical neural interfaces. (a) Direct illumination systems with no scanning elements provides non-specific illumination. (b) Scanned systems where lateral (XY) and varifocal (Z) elements provide 3D positioning of a spot of light to perform sequential light delivery to individual cells. (c) Holographic systems where a spatial light modulator is configured to project light in parallel to multiple neurons, with single-cell precision.

Fig. 1 shows the three main approaches to light delivery into neural tissue, namely (1) direct delivery of broad static illumination, such as an LED or optical fiber delivering light to a population of neurons, (2) scanning methods, in which a single spot of light (either diffraction limited or matched to the dimensions of the neuron's soma) is sequentially placed onto

C. Yalcin, N. T. Ersaro, M. M. Ghanbari, G. Bocchetti, S. Faraji Alamouti, L. Waller, R. Muller are with the Department of Electrical Engineering and Computer Sciences, University of California, Berkeley, Berkeley, CA USA 94720.

N. Antipa is with the department of Electrical and Computer Engineering in University of California San Diego, La Jolla, CA USA 92093.

D. Lopez is with the School of Electrical Engineering and Computer Science in Pennsylvania State University, State College, PA USA 16801.

N. C. Pégard is with the department of Applied Physical Sciences in University of North Carolina at Chapel Hill, Chapel Hill, NC USA 27599.

L. Waller and R. Muller are with Chan-Zuckerberg Biohub, San Francisco, CA USA 94158.

target neurons by the use of lateral scanners (XY scanning) and varifocal elements (Z scanning), and (3) holography, in which the stimulation or fluorescence excitation pattern is sculpted into a hologram to simultaneously target multiple neurons of interest. While (1) allows for a simple optical system, lack of precise spatio-temporal control over illumination limits the use case of these systems to bulk optogenetics applications in which genetically identical populations of neurons that express the optogenetics encoders are always stimulated simultaneously as a unique ensemble. For imaging applications, broad illumination entirely places the burden of reconstructing the 3D scene on the imaging system, either through a scanner located in the imaging path, or through computational imaging methods wherein the 3D scene is reconstructed from a single 2D image at the cost of higher computation resources [5]. On the other hand, scanning and holographic light delivery systems allow for arbitrary placement of cell-level stimulation features in a millimeter-scale field-of-view (FoV). This capability allows not only precise neuromodulation in individual neurons targeted amongst thousands of photosensitive neurons, but also selective interrogation and fluorescence excitation of different locations in the volume, enabling time-multiplexed readout, and greatly simplifying scene reconstruction, to the point where a single photodetector can serve as the imaging element [6].

A typical example of single neuron targeting in the cerebral cortex involves target sizes of down to 10 µm, within a FoV of 1mm x 1mm (lateral) x 300 µm (axial), using wavelengths that range from 450nm to 1500nm. For the optical system to not be a significant bottleneck to the overall throughput of the system, its components must have refresh rates of at least several kHz, as the settling time of optical elements are added to the exposure time of opsins and GEVIs to determine the overall throughput of the system. For scanned systems, a high optical system refresh rate directly translates to higher throughput as targets have to be addressed sequentially. Speckle noise, which are high spatial frequency artifacts usually encountered in holographic systems, can also be reduced through utilization of high refresh rates. Time averaging of multiple holograms suppresses speckle noise, improving the accuracy of the resulting light distribution as the refresh rate increases beyond the regime in which opsins operate [7].

A variety of optical modulation technologies have previously been employed to achieve dynamic patterning of illumination in target neural tissue volumes. For scanned systems, galvanometric scanner mirrors are commonly used lateral scanners, and can achieve kHz speeds, allowing high-throughput random access operation [8]. In contrast, state-of-the-art varifocal elements are electrically tunable lenses (ETLs) and have settling times that exceed 15 ms, severely bottlenecking the response time of the overall optical system [9]. In another commercially available technology, the liquid crystal (LC) medium, the fluidic settling behavior of the LC molecules limits the refresh rate to <500 Hz, especially for longer wavelength ranges (>800 nm) [10] [11].

Faster optical modulation techniques have also been employed in varifocal applications, but such approaches either lack the crucial capability of random-access scanning or require impractical drivers preventing easy integration into random access all-optical interfaces. One such method is the tunable acoustic gradient index of refraction (TAG) lens, which uses standing acoustic waves in fluidic environments to modulate the local index of refraction, sweeping the focal point of the optical system across a given range [12]. While these devices operate at tens of kHz, their resonant operation prohibits dwell capability. Another method employs continuous deformable mirrors (CDM), which can achieve kHz refresh rates with dwell capability, but require drive voltages on the order of 100V or more to achieve meaningful actuation ranges [13]. This requirement complicates driver requirements, increases system size, and limits the number of independent elements in an array that can be feasibly driven. CDMs also suffer from coupled actuation between neighboring pixels, preventing utilization of phase wrapping in the applied hologram and causing non-idealities, thereby limiting focus tuning range [14]. Digital micromirror devices (DMDs) are a fast and compact alternative that perform binary amplitude modulation, which can produce configurable Fresnel zone plates for varifocal operation. However, these devices suffer from very poor optical efficiency, with <5% of the optical power input to the system making it to the focal point [15].

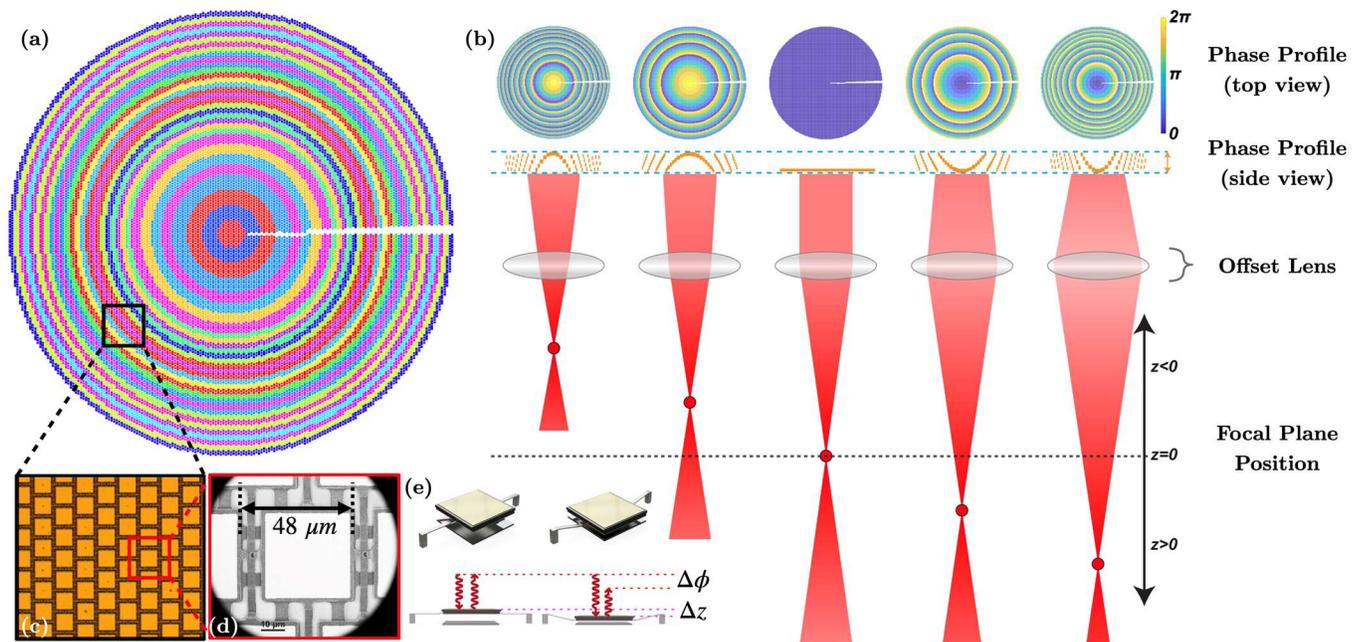

Fig. 2: (a) Wiring scheme of the annular array with 23,852 square-shaped mirrors arranged into 32 individually addressable concentric rings, capable of introducing radially symmetric phase patterns. (b) Example array configurations for tuning the focal point of an offset lens. (c) Close-up photo of the micromirror array. (d) SEM image of a single mirror. (e) Principle of operation of a piston-motion micromirror depicting translation of vertical displacement difference of mirrors to phase difference of reflecting light. Figure adapted from [19].

Spatial light modulators (SLMs) used for holographic systems suffer from technology-specific limitations. SLMs are arrays of phase and/or amplitude modulating elements that can be dynamically configured as the hologram of the desired light intensity distribution in the target volume [16]. State-of-the-art SLMs utilize LC on silicon (LCoS) technology and are limited to <500 Hz like LC lenses [10]. However, piston-motion micromirrors are a promising class of unit elements that offer high-speed operation [17]. In such structures, a segmented planar mirror is vertically displaced at each pixel to alter the travel path of locally incident light, imprinting a phase mask onto an incident coherent wavefront. These structures can operate with time constants on the order of 100 μs or less, offering multiple orders of magnitude of improvement in refresh rate compared to LCoS SLMs. With such high speeds, a random-access all-optical neural interface would become purely opsin-limited for neurostimulation, and optical settling would be on the order of exposure time for GEVI-based fluorescence imaging. Furthermore, a reduced degree-of-freedom SLM can be configured as a spherical phase surface and can serve as the varifocal element in a scanning system [18], while not requiring complicated driving and integration schemes like conventional SLMs. For example, we have previously demonstrated that an annular MEMS mirror array consisting of 23,852 mirrors wired as 32 independently addressable concentric rings can be used for focus tuning [19].

The operating principles of the MEMS-based varifocal mirror are shown in Fig. 2. We designed and fabricated the array using the MEMSCAP PolyMUMPs process with thickness modifications and custom Au lift-off post-processing for metallization. Each micromirror pixel consists of a fixed bottom electrode that, through parallel-plate capacitive transduction, actuates an electrically biased mirror body supported by two clamped-guided suspension beams. Pixel-level phase shifting is achieved as the travel path of incident light is increased by an amount that corresponds to twice the mirror actuation displacement, as depicted in Fig. 2e. The array is capable of introducing radially symmetric phase masks, patterning incident beams into spherical wavefronts, effectively tuning the focal point of the overall optical system.

In order to realize a compact optical scanning system, drive electronics for the MEMS mirrors need to be integrated onto a single IC that can accommodate for process variations and array scale drive requirements. To design this driver, we performed the analysis described in Section II to determine the actuation resolution requirements for three applications: 3D point scanning for single-cell precision, point cloud holography for multi-target optogenetic neurostimulation, and mesh-based holography for the generation of arbitrary shapes, such as light sheets for fluorescence imaging. We identified that 6-bit drive of phase modulators is sufficient to generate high fidelity holograms for all of these approaches. We then designed a driver ASIC, described in Section III, capable of supporting all three applications up to an SLM array size of 200x200, or a varifocal mirror with up to 32 concentric rings. The ASIC features a reconfigurable nonlinear 6-bit DAC that can be programmed to implement the inverse nonlinearity of the MEMS array being driven, correcting global mismatches in MEMS fabrication as well as the inherent nonlinearity of electrostatic actuation. Electrical and optical measurement results are presented in Section IV. Finally, a summary and comparison to the state-of-the-art are presented in Section V.

## II. MEMS MIRROR ACTUATION REQUIREMENTS

To determine the relationship between actuation resolution and hologram quality, SLM performance was simulated across various array formats at a fixed pitch of 22.5 μm. In the simulation, a 4f optical system imaging a laser spot was considered with the SLM located in the Fourier plane and light intensity distribution calculated at the target volume through Fresnel propagation. A focal length of 9mm was used and observation planes were located inside a range of ±1.5mm from the focal plane. Three target light intensity distribution cases were considered: (1) steering a single spot in X, Y and Z for 3D point-scan stimulation, (2) generation of a 3D point cloud for multi-target holographic optogenetics, and (3) generation of arbitrary mesh-based shapes for general purpose holography. Holograms corresponding to target intensity distributions were computed analytically for the single-point scanning case and using global Gerchberg-Saxton algorithm for the multi-point and mesh-based cases. The resulting phase masks were then discretized and summed with random noise to account for finite actuation resolution. Target intensity pattern $T(x, y, z)$ is specified as a binary amplitude pattern with pixel values of 0 or 1. Generated intensity pattern $G(x, y, z)$ is computed through the simulation of light propagation through the 4f optical system with the SLM expressing discretized phase mask. To quantify the quality of the generated pattern, accuracy (α) and efficiency (η) metrics were used [20]. α is a measure of similarity between the desired intensity pattern and the generated intensity pattern, and is computed as the cross-correlation of the two patterns:

$$\alpha = \frac{\sum_{x,y,z} G(x,y,z)T(x,y,z)}{\sqrt{\left[\sum_{x,y,z} G(x,y,z)^2\right]\left[\sum_{x,y,z} T(x,y,z)^2\right]}} \qquad (1)$$

η is a measure of how much of the projected energy is in the targeted voxels, and is calculated using the expression:

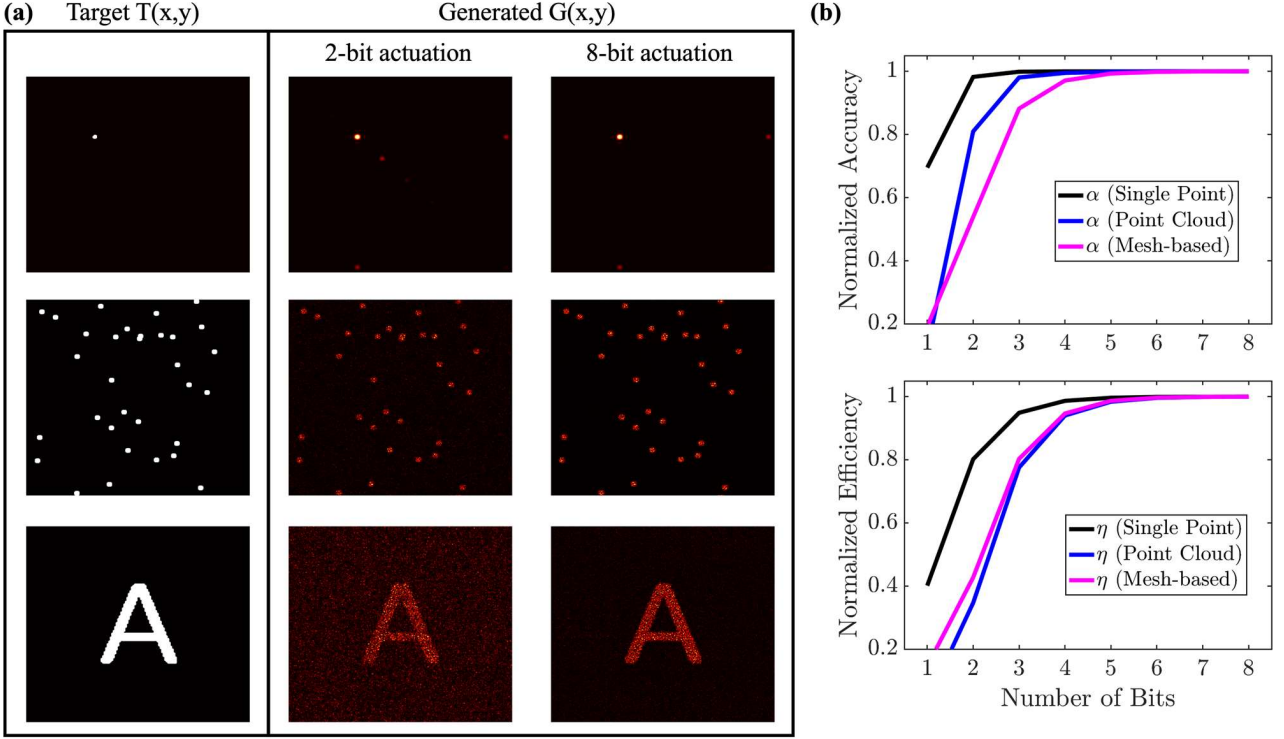

Fig. 3: (a) Examples of target T(x,y|z=-1.5mm) and generated G(x,y|z=-1.5mm) light intensity distribution simulations for single-point scanning, and point cloud and mesh-based approaches of hologram generation, with images shown for 2- and 8-bit actuation resolution cases. At lower resolutions, artifacts such as higher order diffraction modes (top row) or excessive speckle noise (middle and bottom rows) degrade hologram quality. (b) Accuracy (α) and efficiency (η) metrics, normalized to an infinite precision SLM. Results show 6 bits of resolution in phase modulation is sufficient to generate highly accurate and efficient holograms for all approaches.

$$\eta = \frac{\sum_{x,y,z} G(x,y,z)T(x,y,z)}{\sum_{x,y,z} G(x,y,z)} \qquad (2)$$

The metrics defined by (1) and (2) were then normalized to the metric achieved by an SLM of the same array format, with infinite actuation resolution. Fig. 3 shows target and generated images, and normalized α and η for various drive resolutions in three kinds of SLM applications. In all cases, the accuracy of the generated hologram encounters a small amount of degradation at 4 bits and saturates at 6 bits of resolution in phase modulation. Therefore, in this work we have implemented a mirror driver that can provide a 6-bit control in phase modulation.

The piston-type MEMS mirrors used in this work are electrostatically actuated parallel-plate structures. The phase of the incoming beam is modulated through vertical displacement of this structure through a voltage applied across the two electrodes. Fig. 4 shows simulated voltage actuation curves for a sample mirror structure, quantized with 6-bits of actuation, alongside dashed lines representing process corners with 5% thickness variation of the structural layers. This displacement-actuation voltage relation is nonlinear with respect to the applied voltage for a given displacement approximated by the equation [13]:

$$V(\Delta z) = \sqrt{a(b-\Delta z)^2 \Delta z} \qquad (3)$$

where $\Delta z$ is the vertical displacement from the resting height,

and a, b are fitting constants susceptible to process variations causing die-to-die and pixel-to-pixel mismatches, requiring per-part calibration. A conventional solution to this problem is to utilize an array of discrete high-resolution linear DACs and perform calibration using look-up tables. Since V(Δz) is nonlinear, a linear DAC wastes dynamic range in the region of the curve where the transduction gain is low, and hence a higher voltage LSB can be used. Furthermore, discrete DAC arrays significantly increase the size of the system, limiting the potential applications of high actuator count SLMs. A driver ASIC with an integrated voltage generation scheme stands to shrink the system size, allowing for integration of SLMs into compact holography systems, such as optogenetic stimulation devices for moving animals.

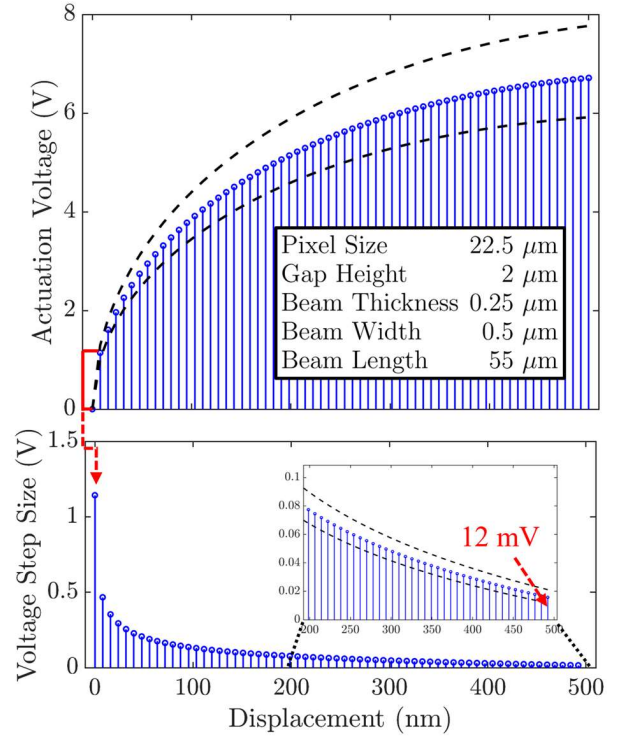

| Pixel Size | 22.5 $\mu m$ |
| Gap Height | 2 $\mu m$ |
| Beam Thickness | 0.25 $\mu m$ |
| Beam Width | 0.5 $\mu m$ |
| Beam Length | 55 $\mu m$ |

Fig. 4: Voltage vs. displacement curve for a simulated MEMS micromirror with dimensions provided in the table. Dashed lines represent process corners with ±5% thickness variation.

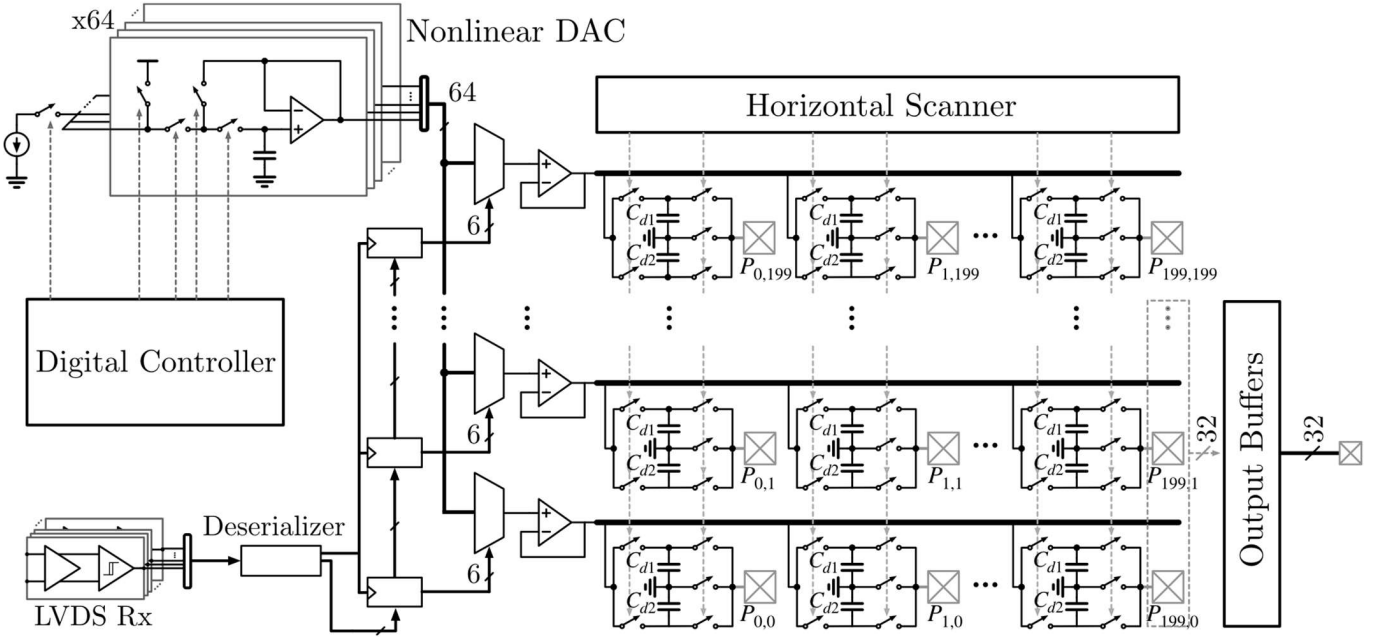

Fig. 5: Simplified block diagram of the mirror driver ASIC.

## III. DRIVER ASIC IMPLEMENTATION

To overcome both the global variations in the MEMS process, and to provide a linear digital code-to-displacement conversion, we have developed a driver ASIC that employs a reconfigurable nonlinear 6-bit DAC [21]. Electrical connection to MEMS devices can be established either through 5.4x5.4 $\mu m^2$ pad openings arranged in a 200x200 pixel array for fully independent SLM operation, or through 32 wire-bond pads for low degree-of-freedom MEMS arrays. To minimize power consumption while retaining the required actuation range for MEMS devices with >0.5 $\mu m$ lateral features, the ASIC was designed with 8 V drive capability. As shown in Fig 4, for linearly spaced 64 displacement levels, the voltage differences between adjacent codes range from 1.1 V in the lowest end to 12 mV in the highest end across process corners for a simulated MEMS device with 500 nm vertical displacement under 0-8 V drive. The drive circuit for such an actuator requires 11-bit accuracy in the higher actuation regime, while only requiring 4-bit accuracy in the lower end of the curve. This property was exploited by designing a reconfigurable nonlinear DAC that reuses its precision setting capacitors as sample & hold capacitors to save power and area compared to a linear DAC

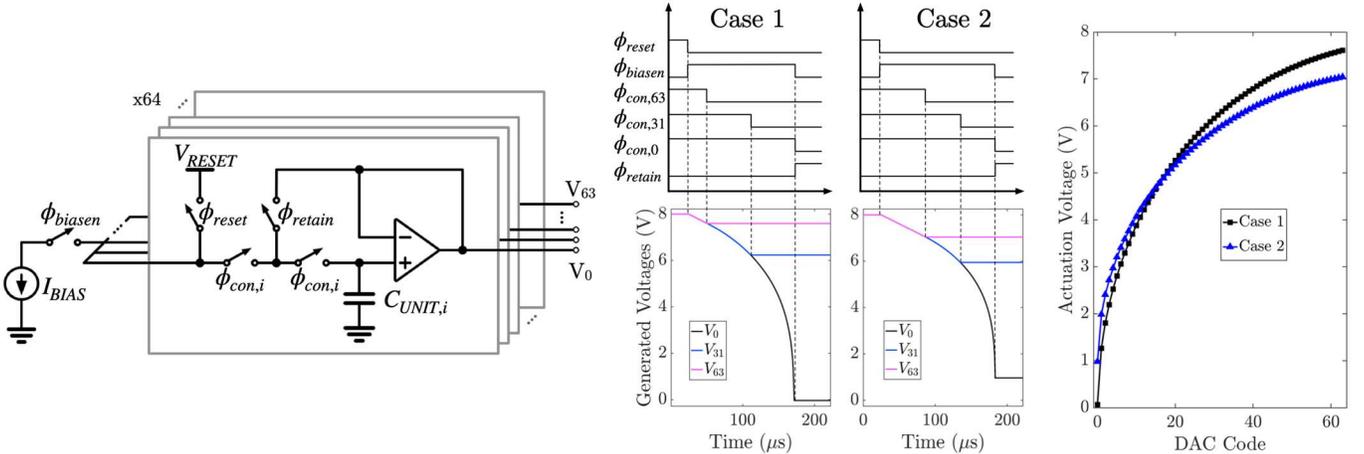

Fig. 6: Generation of nonlinearly spaced voltages that correspond to linearly spaced displacement levels of the mirrors. Measured results of cases that correspond to two MEMS samples are shown on the right.

that spans the entire dynamic range.

Fig. 5 shows the simplified block diagram of the ASIC. The nonlinear DAC generates 64 voltages that correspond to linearly spaced mirror displacement levels. Mirror displacement data is transmitted via a 4 Gbps LVDS link consisting of four channels, operating at 1 Gbps/channel with 6b/8b encoding to ensure DC balance. This data is then scanned into a shift register chain to configure analog multiplexers and select the corresponding voltages to be written to each pixel's DRAM cell. Each unit pixel contains a pad opening to bond a MEMS mirror, and two capacitors that comprise two DRAM cells. 32 of these pixels are connected to output buffers to drive the internal voltages off-chip. The entire array has a refresh rate of 10 kHz, although it is possible to window only the 32 pixels driving the output buffers to achieve refresh rates up to 2 MHz.

The nonlinear 6-bit DAC is composed of two sections: a voltage reference to generate and retain the 64 analog voltage values that correspond to each level of vertical displacement for a given actuation curve, and a distributed analog multiplexer and buffer pair per row to select and write the corresponding voltage to each pixel. Fig. 6 shows the schematic of the reference voltage generator section, alongside timing diagram with the generation and retention of voltage levels for two possible nonlinear actuation curves. A capacitor bank containing 64 unit capacitors ($C_{UNIT}$=2.2pF), a current source for controlled discharge, and a reset switch are all connected to a common node. Initially, all capacitors are reset to $V_{RESET}$=8V, and then discharged through the current source ($I_{BIAS}$=2μA). Capacitors are sequentially disconnected from the common node to sample voltages that correspond to their respective codes through the timing of $\phi_{con.i}$ signals. Timing is controlled by a state machine and on-chip memory containing discharge times for each code (8 bits/code) that define how many periods of $T_{CLK}$ (50ns) discharge should occur, to yield $\Delta T_i$. The generated voltage for a given code $i$ is

$$V_i = V_{i+1} - \frac{I_{BIAS} \times \Delta T_i}{C_{TOT}(i)} \qquad (4)$$

where $C_{TOT}(i)$ is the total capacitance connected to the discharge node for code $i$. As capacitors are removed from the common node, discharge speeds up and precision of the generated voltage decreases. Programmability of this voltage generation scheme allows for simultaneous cancellation of mirror nonlinearity and calibration of process variations. Voltages are buffered with rail-to-rail class AB amplifiers and distributed to the rest of the ASIC, to serve as reference voltages in the DRAM write chain depicted in Fig. 6. Due to the leakage of stored charge on the capacitors to the bulk of the switch devices, the nonlinear DAC is refreshed every 2.5 ms, keeping drift <0.5 LSB error in mirror position. With typical values of discharge current and discharge durations, refresh operation takes <200 μs. While the DAC refresh is a periodic event, discharge durations are calibrated once per MEMS device and programmed into the ASIC during startup.

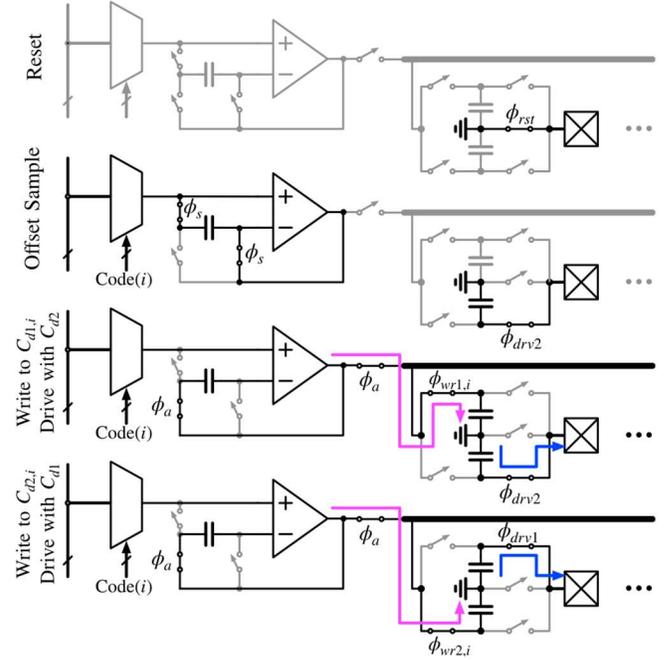

Fig. 7: Principle of operation of the DRAM write chain with 4 phases of configuration shown. The two pixel capacitors are utilized in a ping-pong fashion, enabling global shutter operation to minimize down time between frames.

The pixels for array-scale drive are laid out in a 200x200 grid at a pitch of 22.5 μm and with 5.4 μm x 5.4 μm pad openings for per-pixel MEMS connection. Each pixel contains five switches and two MOM capacitors ($C_{d1,i}$ and $C_{d2,i}$, 250 fF each) that serve as analog DRAM elements. The flow of operation to update the drive voltages in the pixel array is shown in Fig. 7. Digital select codes are transferred to the chip through the

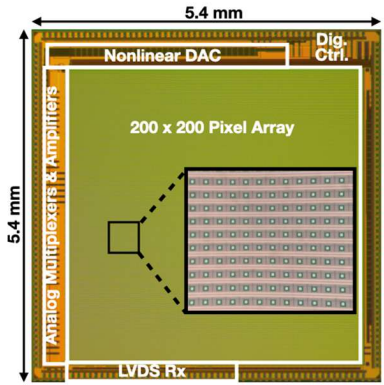

| Driver ASIC Specifications | |
|---|---|
| | This work |
| Technology | 40nm + HV |
| Refresh Rate | 10kHz - 2MHz* |
| Pixel Array Size | 200×200 |
| Pixel Pitch | 22.5 μm |
| Drive Voltage Range | 0-8 V |
| Actuation Resolution | 6 bits |
| Power Consumption | 308 mW |

* 10 kHz for the entire array, 2 MHz for the 32 buffered outputs

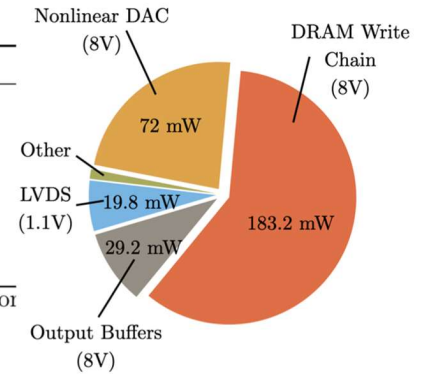

Fig. 8: Chip micrograph with the inset showing MEMS pad openings, chip specifications and power breakdown.

LVDS link and distributed to each row through a chain of shift registers. For each write operation, the MEMS capacitor is reset to VSS to prevent frame-to-frame hysteresis, the corresponding reference voltage is selected, the offset of the amplifier is cancelled through an auto-zero phase, and the buffered value is written to the corresponding pixel. The two DRAM capacitors in the pixel operate in a ping-pong fashion, alternating between storing value for the next frame and driving the MEMS pad. The capacitors switch roles with each new frame to provide global-shutter operation, minimizing downtime between subsequent frames and eliminating rolling shutter artifacts, which would prolong the effective settling time of the optical element. As the simulated value of the parallel plate capacitance of the mirror structure is <10 fF, there is negligible charge sharing between the pixel capacitance and the actuator, which is accounted for by pre-distorting the reference voltages.

## IV. MEASUREMENT RESULTS

The IC was fabricated in TSMC's 40 nm HV CMOS technology node. The die micrograph and power consumption breakdown are shown in Fig. 8. Measurements are divided into two sections: electrical measurements of the ASIC to verify the performance metrics of the nonlinear DAC and DRAM write chain, and optical measurements taken driving 32-channel MEMS varifocal mirror [19] to demonstrate optical functionality and characterize precision and speed of electromechanical actuation of the ASIC-MEMS system.

### A. Electrical Measurements

The nonlinear DAC was first characterized separately from the MEMS to verify that the electrical performance meets application specifications. Importantly, the ASIC should not cause more than 1 LSB error in displacement for any supported MEMS mirror actuation curve, including the drift caused by leakage from the DAC storage capacitors discussed in the previous section, which was budgeted 0.5 LSB, leaving another 0.5 LSB for the rest of the write chain. To determine the edge constraints, two extreme mirror actuation cases were considered: (1) a highly nonlinear voltage-displacement response such as the mirror model presented in Fig. 4, and (2) a 0-8 V fully linear voltage-displacement response that is more pessimistic than any real actuation curve would be in the lower code regime. These two constraints are stringent on opposite ends of the actuation range. Fig. 9 shows a comparison between the two sets of specifications, (1) indicated by magenta and (2) indicated by the green dashed lines, together with the measured post-calibration precision and maximum residual error of the nonlinear DAC for each code. Here, the precision is defined as the refresh-to-refresh standard deviation of the voltage corresponding to each code, and results from the noise of the DAC current source and amplifiers in the write chain. The maximum residual error refers to the change that can be induced in the mean output voltage by tuning the discharge time of the given code by 1 bit, and is limited by the clock period, discharge current of the DAC and the code capacitance as described in Equation 4. This value represents how close a given code is guaranteed to approach an arbitrary voltage. The results show that the joint error in mirror displacement due to residual error and finite precision of the DAC is <1 LSB for a wide range of possible mirror actuation profiles.

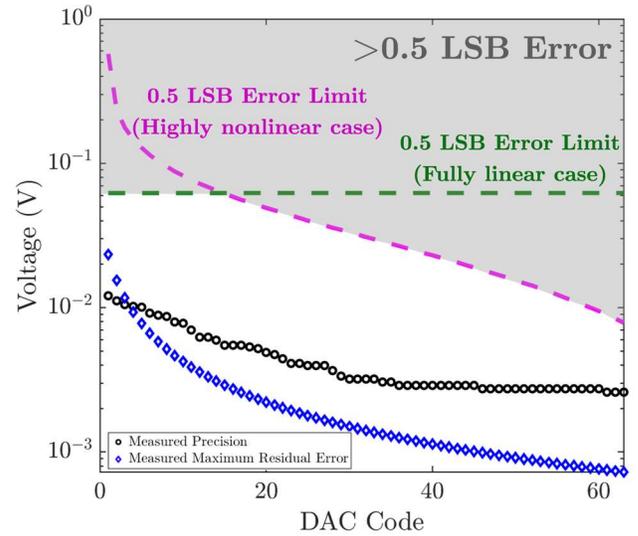

Fig. 9: Measured precision and maximum residual error of the nonlinear DAC vs. DAC code. Two sets of constraints are also shown in dashed lines that correspond to the most stringent cases for different ends of the actuation curve. Maximum residual error of the DAC is defined as the change that is induced in mean DAC output for a given code when the code discharge duration is changed by 1 bit. Precision of the DAC is the standard deviation of a code output voltage measured refresh-to-refresh.

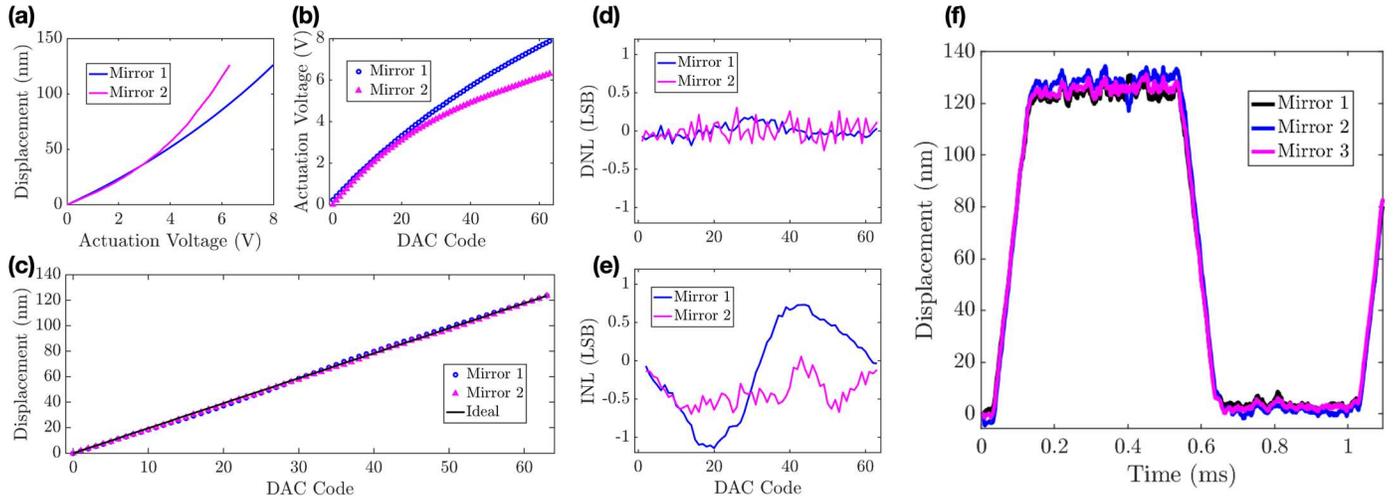

Fig. 10: Static and dynamic measurements of the ASIC-MEMS system performed under a DHM. (a) Measured displacement vs. voltage behavior for two mirrors. (b) Measured transfer curve of the nonlinear DAC post-calibration for the two mirrors. (c) Measured displacement vs DAC code behavior for the two mirrors. (d-e) DNL and INL of displacement vs. DAC code. (f) Dynamic behavior of three mirrors while the ASIC was configured to switch between the two extreme ends of the actuation curve at 2 kHz.

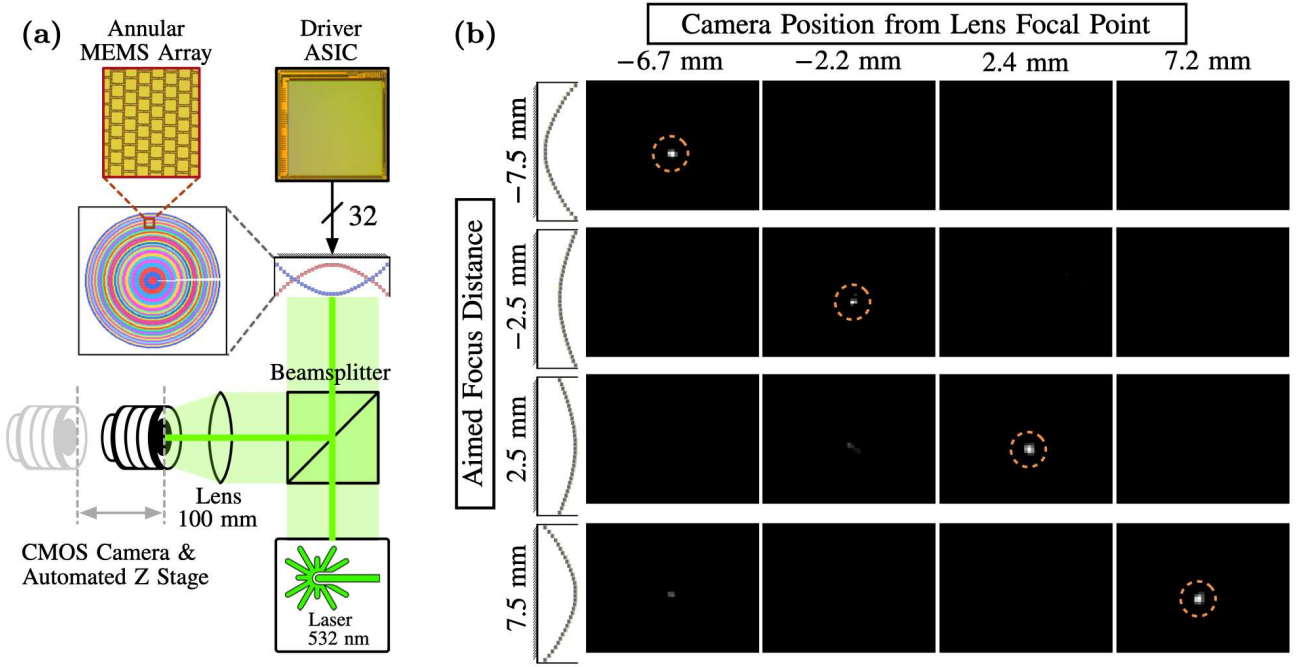

Fig. 11: (a) Optical measurement setup for the tunable lens system formed by the ASIC and 32-channel MEMS array. During the measurements, ASIC was programmed to implement the inverse nonlinearity of the mean actuation curve for the entire array. Effect of local mismatches are mitigated by the highly redundant nature of the radially symmetric phase masks being used. (b) Z-stack measurements relative to background illumination for four target focus depth configurations.

### B. Optical Measurements

The 32-channel annular MEMS array was driven with the ASIC to form the varifocal system. A digital holographic microscope (DHM) was used to observe the behavior of individual mirrors inside the array. Since the MEMS array used in this work has a full-scale drive range of 32V, a -20V bias voltage was applied to the top electrode of each mirror to operate the device inside the high transduction gain region of the actuation curve.

Static measurements of two individual mirrors were performed to generate DNL and INL characteristics of the digital code-to-displacement conversion, and the results are shown in Fig. 10a-e. First, the mirror actuation curves were extracted using a discrete 14-bit linear DAC. The ASIC was then programmed to implement the inverse nonlinearity of the mirror under study. For each digital input code, displacement value after full mechanical settling was recorded. The process was repeated for a mirror from a different MEMS die. Maximum DNL and INL values recorded across all codes and both mirrors were 0.21 LSB and 1.14 LSB, respectively.

Dynamic measurements were made with the stroboscopic mode of the DHM and three mirrors on the same die were simultaneously observed while being driven between two displacement values. The results are shown in Fig. 10f and the maximum 10-90% rise/fall times for these mirrors were measured to be 80 μs and 82 μs, respectively.

To demonstrate optical utility of the ASIC-MEMS system, a 4f imaging system was constructed to image a laser point, with

Table 1: Varifocal Element Comparison Table

|  | [9] | [11] | [12] | [13] | [15] | **This work** |
|---|---|---|---|---|---|---|
| Actuator Type | ETL | LC Lens | TAG Lens | CDM | DMD | **Piston Mirror** |
| Settling Time* | > 15 ms | > 3 ms | < 15μs | < 100μs | 45μs | **82 μs** |
| Dwelling Capability | Yes | Yes | No | Yes | Yes | **Yes** |
| Required Driver Voltage | < 20V | 10 − 100V | < 20V | 250V | 1V | **8V** |
| Volumetric Efficiency** | N/A† | >90% | N/A† | – | 3.7% | **38%** |
| Number of Actuators | – | – | – | 1k-3k | 786k | **24k** |
| Number of Independent Channels | – | – | – | 1k-3k | 786k | **32** |
| IC Power Consumption | – | – | – | – | 2-4.4 W | **308 mW** |
| Required Datarate | – | – | – | – | 25.6 Gbps | **3 Mbps** |

* 10% to 90% settling time
** Measured at the focal plane
† Not a diffractive device

the annular MEMS array located at the Fourier plane. A CMOS camera on an automated z-stage was used to capture images formed in the target volume for various configurations of the tunable lens. Fig. 11 shows the diagram of the optical setup and images taken at 4 depths for 4 curvature configurations of the varifocal mirror. While deviations from aimed focus depths were observed due to imperfect alignment of the optical system, this is a deterministic effect that can be corrected by a lookup calibration of aimed depths vs. observed focal plane depths. The volumetric efficiency of the system was quantified as the ratio of the energy located inside the spot full width at half maximum (FWHM) to the total energy located in the field of view and was found to be 38% at the focal plane of the lens. The spot FWHM was measured to be 10 μm in X and Y directions, and 900 μm in the axial direction with a full-scale continuous tuning range of ±10mm when used with a f=100 mm lens, spanning 22 fully resolvable depth planes at refresh rates greater than 12 kHz. Through the demagnification of the imaged spot, this device can address 10 μm-sized targets across an axial range of 220 μm.

## V. Summary and Discussion

We present a varifocal mirror system for high-speed, random-access 3D point-scanning systems for optogenetic stimulation. The system is comprised of an annular array of piston-motion MEMS mirrors wired into 32 concentric rings, and a driver ASIC. The ASIC features a reconfigurable nonlinear DAC that provides a linear code-to-displacement conversion by correcting the inherent nonlinearity of electrostatic actuation as well as global MEMS process variations. The system can address 22 distinct depth planes with refresh rates >12 kHz. Our system's refresh rate exceeds the two most common varifocal elements (ETLs and LC lenses) by a factor of >36x, possesses random-access and dwelling capability lacking in resonant devices such as TAG lenses, and requires only 8V drive allowing scalability to large array formats. Compared to DMD-based approaches, this work offers 10x higher volumetric efficiency and 10x lower power consumption, using 33x fewer actuators.

An array of micromirrors with pixel-level independent actuation through the ASIC could unify lateral scanning and varifocal operation in a single chip-scale device, significantly miniaturizing 3D point scan systems. For example, a 10 kHz, 200x200 pixel SLM that can be supported by the ASIC in this work could target hundreds of neurons in a 500x500x500 μm³ volume of brain within 1 ms, a relevant timescale for neural signaling that corresponds to the duration of a single action potential.

Such a high speed SLM can also be extended to applications outside of neuroscience, such as 3D holographic near-eye displays for AR/VR systems by overcoming two attributes that are limited by the slow refresh rates of the LCoS SLMS used in current systems. A higher refresh rate allow time multiplexing between three color domains to enable full-color holographic displays using a single SLM. Simultaneously, the time averaging capability enabled by the excess frame rate can be utilized through existing speckle noise reduction techniques to improve hologram accuracy and overall image quality.

## Acknowledgements

The authors thank the sponsors of Berkeley Wireless Research Center (BWRC), the TSMC University Shuttle Program for chip fabrication, and Prof. Ming Wu. This work was funded by Chan Zuckerberg Biohub and the McKnight Technological Innovations in Neuroscience Award.